\documentclass{article}
\usepackage{arxiv}
\usepackage[utf8]{inputenc} %
\usepackage[T1]{fontenc}    %
\usepackage{hyperref}       %
\usepackage{url}            %
\usepackage{booktabs}       %
\usepackage{amsfonts}       %
\usepackage{nicefrac}       %
\usepackage{microtype}      %
\usepackage{graphicx}
\usepackage[numbers]{natbib}

\def\email#1{\texttt{#1}}

\usepackage{caption}
\usepackage{xcolor}
\usepackage{url}
\usepackage{hyperref}
\usepackage{multicol}
\usepackage{multirow}
\usepackage{subcaption}
\usepackage{amsfonts}
\usepackage{paralist}
\usepackage{todonotes}
\usepackage{booktabs}
\usepackage{makecell}

\newif\ifdraft
\draftfalse

\newif\ifrevised
\revisedfalse

\newif\ifblind
\blindfalse

\newcommand{\revision}[2]{
    \ifrevised
    \textcolor{red}{#2}\todo[color=red!30]{Re: #1}
    \else
    {#2}
    \fi
}

\newcommand{\ttitle}{A Set of Distinct Facial Traits Learned by Machines Is Not Predictive of Appearance Bias in the Wild}
\newcommand{\stitle}{A Set of Distinct Facial Traits Is Not Predictive of Appearance Bias}

\newcommand{\instituteA}{
    Heinz College of Information Systems \& Public Policy\\
    Carnegie Mellon University \\
    Pittsburgh, PA, USA\\
    \email{ryansteed@cmu.edu}
}
\newcommand{\instituteB}{
    Department of Computer Science\\
    George Washington University\\
    Washington, DC, USA\\
    \email{aylin@gwu.edu}
}

\newcommand{\github}{\ifblind \url{https://github.com/anonymous/repo} \else \url{https://github.com/ryansteed/learning-appearance-bias} \fi}

\title{\ttitle}
\author{ 
    Ryan Steed\\
    \instituteA\\
    \And
    Aylin Caliskan\\
    \instituteB\\
}

\date{January 12, 2021}

\renewcommand{\shorttitle}{\stitle}

\draftfalse

\hypersetup{
pdftitle={\ttitle},
pdfauthor={Ryan Steed, Aylin Caliskan}
}

\begin{document}
\maketitle

\begin{abstract}
Research in social psychology has shown that people's biased, subjective judgments about another's personality based solely on their appearance are not predictive of their actual personality traits. But researchers and companies often utilize computer vision models to predict similarly subjective personality attributes such as ``employability." We seek to determine whether state-of-the-art, black box face processing technology can learn human-like appearance biases. With features extracted with FaceNet, a widely used face recognition framework, we train a transfer learning model on human subjects' first impressions of personality traits in other faces as measured by social psychologists. We find that features extracted with FaceNet can be used to predict human appearance bias scores for deliberately manipulated faces but not for randomly generated faces scored by humans. Additionally, in contrast to work with human biases in social psychology, the model does not find a significant signal correlating politicians' vote shares with perceived competence bias. With Local Interpretable Model-Agnostic Explanations (LIME), we provide several explanations for this discrepancy. Our results suggest that some signals of appearance bias documented in social psychology are not embedded by the machine learning techniques we investigate. We shed light on the ways in which appearance bias could be embedded in face processing technology and cast further doubt on the practice of predicting subjective traits based on appearances.
\end{abstract}

\section{Introduction}
\label{sec:introduction}

Researchers and the public have raised concerns about the use of face detection, face recognition, and other facial processing technology (FPT)\footnote{Following \citet{raji2020saving}, we use the term facial processing technology (FPT) to refer to a broad class of applications that rely on representations of individual's facial characteristics, including face detection, face analysis, and face recognition.} in applications such as police surveillance and job candidate screening due to its potential for bias and social harm \citep{raji2020saving, Buolamwini2018Opinion:Skin, Nagpal2019DeepPrejudiced, Snow2018AmazonsMugshots}.\footnote{In response to these results and major activist efforts, Amazon, Microsoft, and IBM recently placed temporary moratoriums on some FPT products for government surveillance \citep{Hao2020TheReview}.} For example, HireVue's automated recruiting technology uses candidate's appearance and facial expression to judge their fitness for employment \citep{Harwell2019AJob}, and the company 8 and Above uses video interviews to construct candidate ``blueprints," which include estimated personality traits such as openness, warmness, and enthusiasm \citep{Raghavan2020MitigatingPractices}. If a surveillance or hiring algorithm learns subjective human biases from training data, it may systematically discriminate against individuals with certain facial features. We investigate whether industry-standard face recognition algorithms can learn to make biased, stereotypical trait judgments about faces based on human participants' perception of personality traits from faces. Quick trait inferences should not affect important, deliberate decisions \citep{Willis2006FirstImpressions}, but humans do display first impression trait biases \citep{Todorov2017FaceImpressions} and those inferences could affect human judgments of other subjective traits like ``employability" or ``attractiveness" that algorithms are actively designed to mimic \citep{Harwell2019AJob}. If off-the-shelf FPT can learn biased trait inferences from faces and their labels, then application domains using FPT to make decisions are at risk of propagating harmful prejudices.

Because the predictions made by machine learning models depend on both the training data and the annotations used to label them, systematic biases in either source of data could result in biased predictions. For instance, a dataset on employment information designed to predict which job candidates will be successful in the future might contain data regarding mainly European American men. If such a dataset reflects historical injustices, it is likely to unfairly disadvantage African American job candidates. Moreover, annotators could introduce human bias to the dataset by labeling items according to their implicit biases. If annotators for a computer vision task are presented with a photo of two employees, they might label a woman as the employee and the man standing next to her as the employer or boss. Such embedded implicit or sociocultural bias leads to biased and potentially prejudiced outcomes in decision making systems. 

In computer vision, models used in face detection or self-driving cars have been proven biased against genders and races \citep{Buolamwini2018GenderClassification, Wilson2019PredictiveDetection}. Some examples of racial and gender biases include gender classifications made by automated captioning systems and contextual cues used incorrectly by visual question answering systems \citep{Hendricks2018WomenModels, Zhao2017MenConstraints, Manjunatha2019ExplicitModels}. These algorithms are actively used in self-driving cars \citep{Geiger2012AreSuite}, surveillance \citep{Ko2008AApplications}, anomaly detection \citep{Mahadevan2010AnomalyScenes}, military drones \citep{Nex2014UAVReview}, and cancer detection \citep{Bejnordi2017DiagnosticCancer}. But while many biases are explicit and easily detected with error analysis, some ``implicit" biases are consciously disavowed and are much more difficult to measure and counteract. Often, these biases take effect a split second after perception in human judgment. These biases are often quantified by implicit association tests \citep{Greenwald1998MeasuringTest, Greenwald2009UnderstandingValidity.}. Computer vision models do embed historical racial or gender biases, but can they also embed these first-impression appearance biases documented in social psychology \citep{Todorov2013ValidationFacial}?

In this study, we investigate whether biases formed during the first impression of a human face can be learned by industry-standard face recognition models. Like implicit biases, ``first impression" appearance biases are split-second trait inferences drawn from other people's facial structure and expression \citep{Willis2006FirstImpressions, Todorov2017FaceImpressions}. \citet{Todorov2005InferencesOutcomes} characterize first impression bias as unreflective and sometimes unconscious. We consider six types of subjective personality trait inferences drawn from faces, each measured in controlled laboratory experiments \citep{Willis2006FirstImpressions, Todorov2013ValidationFacial}: attractiveness, competence, extroversion, dominance, likeability, and trustworthiness.\footnote{These trait inferences are neither objective nor reified. We examine people's subjective perception of personality traits in other face, which amount to biases and prejudices. Real-world outcomes associated with these subjective trait ratings are the result of social biases, not objective ability or personality.} In a rational world, these physiognomic stereotypes \citep{Hassin2000FacingPhysiognomy}, may seem unlikely to influence deliberate decisions, but appearance biases have been shown to predict numerous external outcomes, including election results \citep{Ballew2007PredictingJudgments, Todorov2005InferencesOutcomes}, income \citep{Hamermesh1994BeautyMarket}, economic decisions \citep{Rezlescu2012UnfakeableBehavior, vantWout2008FriendDecision-making}, and military rank \citep{Mueller1996FacialRank}. Notably, appearance bias is not known to be predictive of any objective measure of ability, performance, or personality \citep{Todorov2005InferencesOutcomes}, and empirically they are often wrong about the people they stereotype \citep{Zebrowitz1998BrightEffects., Keating1999PresidentialPerceptions, Mueller1996FacialRank}.

Despite the fact that appearance biases are neither causally linked to nor predictive of actual personality traits, researchers have built machine learning models to predict appearance bias from faces \citep{Yang2017PredictionLSTM, Safra2020TrackingPaintings}. Likewise, HireVue and other companies still advertise predictive models for other subjective attributes such as ``employability" trained on historical data with historical biases \citep{Harwell2019AJob}. We seek to determine whether general, industry-standard face representations can be used to accurately predict subjective, human trait inferences. If so, then face processing technology is at risk of propagating trait inferences embedded in labeled training data. If not, the practice of predicting subject trait inferences and other related personality attributes is even more dubious.

In this paper, we make several contributions towards understanding appearance bias in FPT. First, we design a transfer learning method for extracting general-purpose face representations suitable for state-of-the-art face processing applications. Second, we train our model on computer-generated faces manipulated to display certain personality traits, including not only Caucasian but also Asian and Black faces. Third, we find that while our model is quite good at predicting perceived trait scores for faces produced by \citet{Todorov2013ValidationFacial}'s computational model of appearance bias, it fails to consistently predict perceived trait scores for randomly generated faces. Additionally, while it has been shown that the human perceptions of the competence of political candidates are correlated with election outcomes \citep{Todorov2005InferencesOutcomes, Ballew2007PredictingJudgments}, our model's competence scores do not achieve the same predictive validity. Our experimental results and additional interpretability analysis suggest that generalized representations for face recognition are not suitable for learning subjective biases.

\subsection{Related Work}
\label{sec:related}

There is a wealth of literature measuring the stereotypes perpetuated by image classifiers and other machine learning models, from search results to automated captioning \citep{Kay2015UnequalOccupations, Hendricks2018WomenModels, Kleinberg2017HumanPredictions}. Previous applications of unsupervised machine learning methods demonstrated the existence of social and cultural biases embedded in the statistical properties of language, but little research has been conducted with respect to the biases in transfer learning models for faces or people and even less attention has been paid to the intersection of machine learning and first appearance bias \citep{Caliskan2017, Torralba2011UnbiasedBias}. \citet{JacquesJunior2018FirstAnalysis} review the use of computer vision to anticipate personality traits. \citet{Yang2017PredictionLSTM} use a novel long-short term memory (LSTM) approach to predict first impressions of the Big Five personality traits after 15 seconds of exposure to various facial expressions. 

Most notably, \citet{Safra2020TrackingPaintings} also train a model on a subset of the computer-generated faces produced by \citet{Todorov2013ValidationFacial}.
They learn to predict subjective trustworthiness ratings with facial action units, or facial configurations such as smiling and frowning, and use their model to analyze the evolution of trustworthiness in portraiture. The authors claim that their model can be used to predict trustworthiness for selfies and historical portraits, but the correlation between their model's predictions and subjective ratings from human annotators in external datasets is low. We clarify their results with several modifications: first, we train on Black and Asian faces, in addition to Caucasian faces; second, we use transfer learning to obtain more generalized face representations; and third, we extract representations of the entire face to capture biases related to face structure and color, not just facial actions such as smiling and frowning.

There is a serious concern that face recognition and face modeling techniques may propagate cognitive and historical biases entrenched in human annotations and model design. Our study investigates part of this concern: we evaluate whether first impression appearance biases can be learned with off-the-shelf face processing technology. Can we observe the same biased effects in real-world datasets with a predictive model?

\section{Data}
\label{sec:dataset}

\begin{figure}[!t]
    \centering
    \includegraphics[width=0.2\linewidth, height=0.2\linewidth]{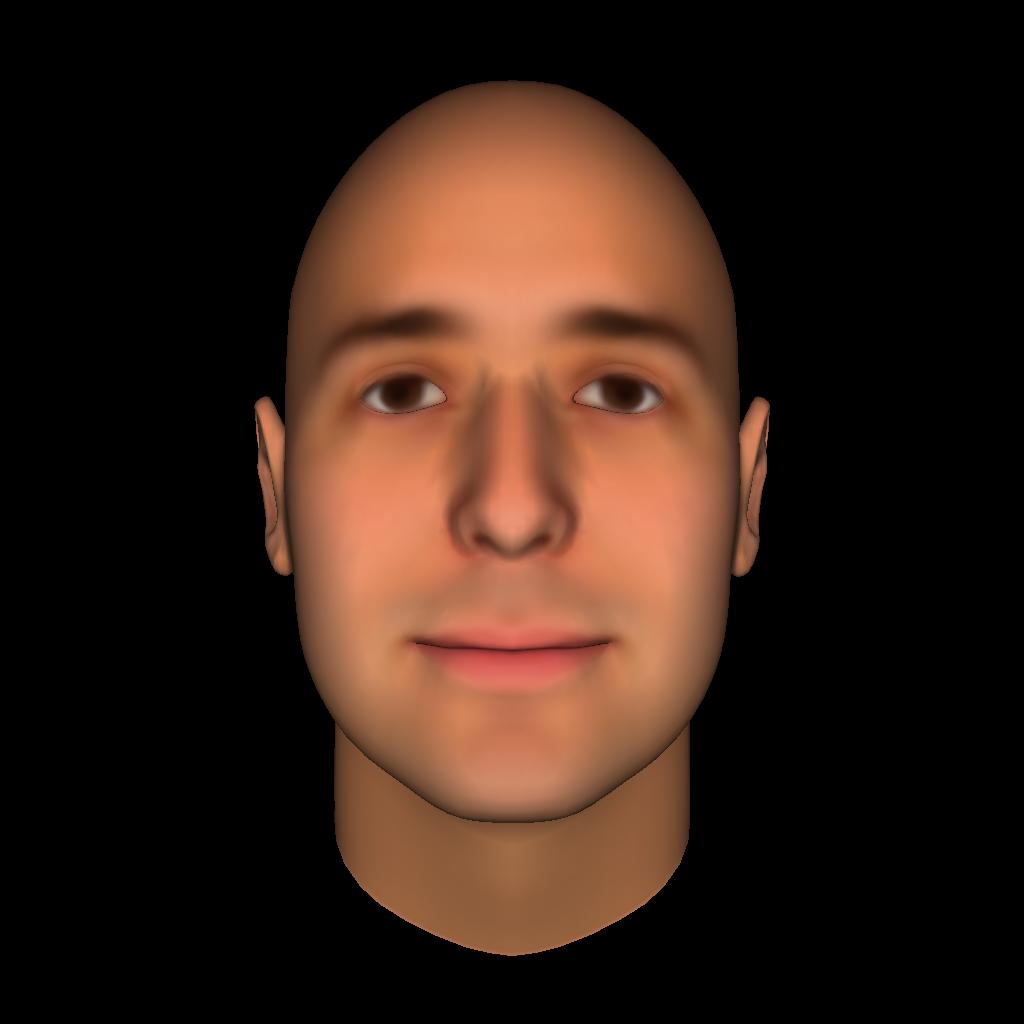}
    \includegraphics[width=0.2\linewidth, height=0.2\linewidth]{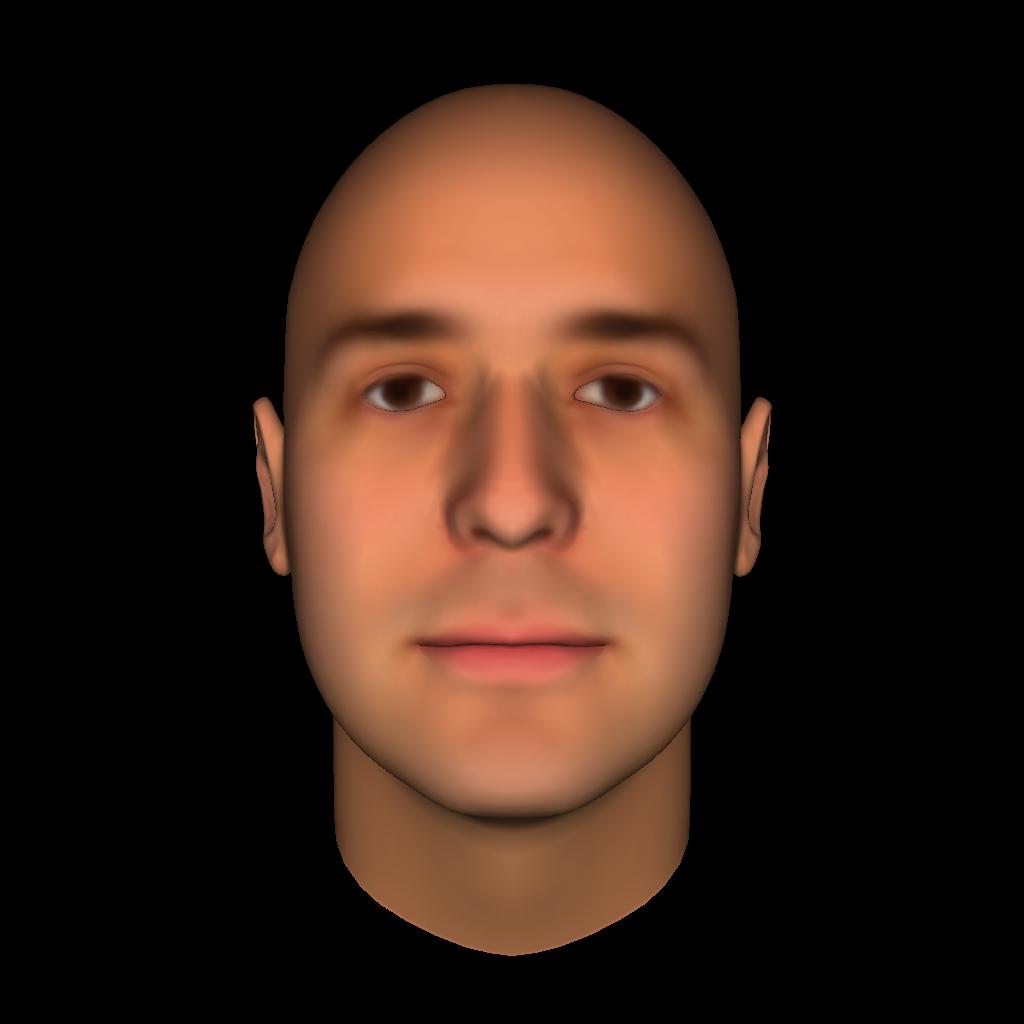}
    \includegraphics[width=0.2\linewidth, height=0.2\linewidth]{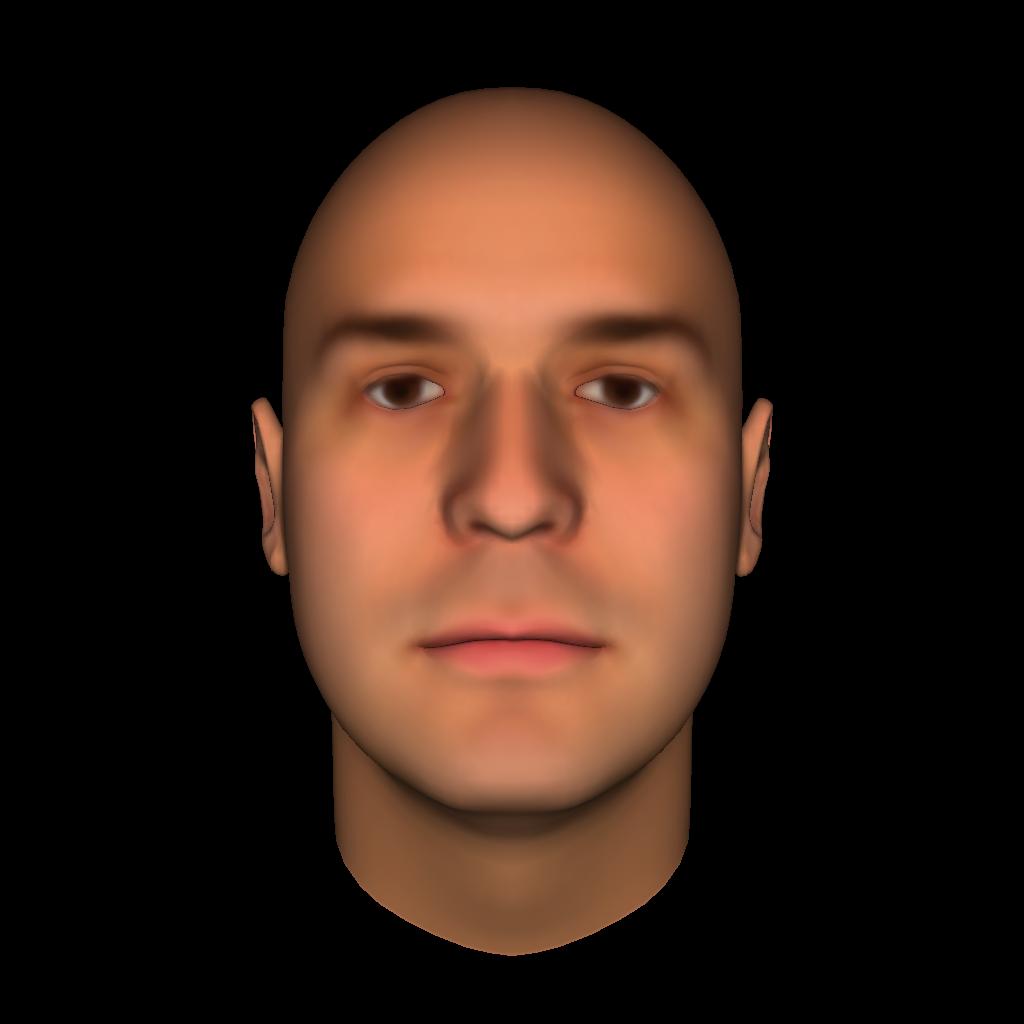}
    \\
    \includegraphics[width=0.2\linewidth, height=0.2\linewidth]{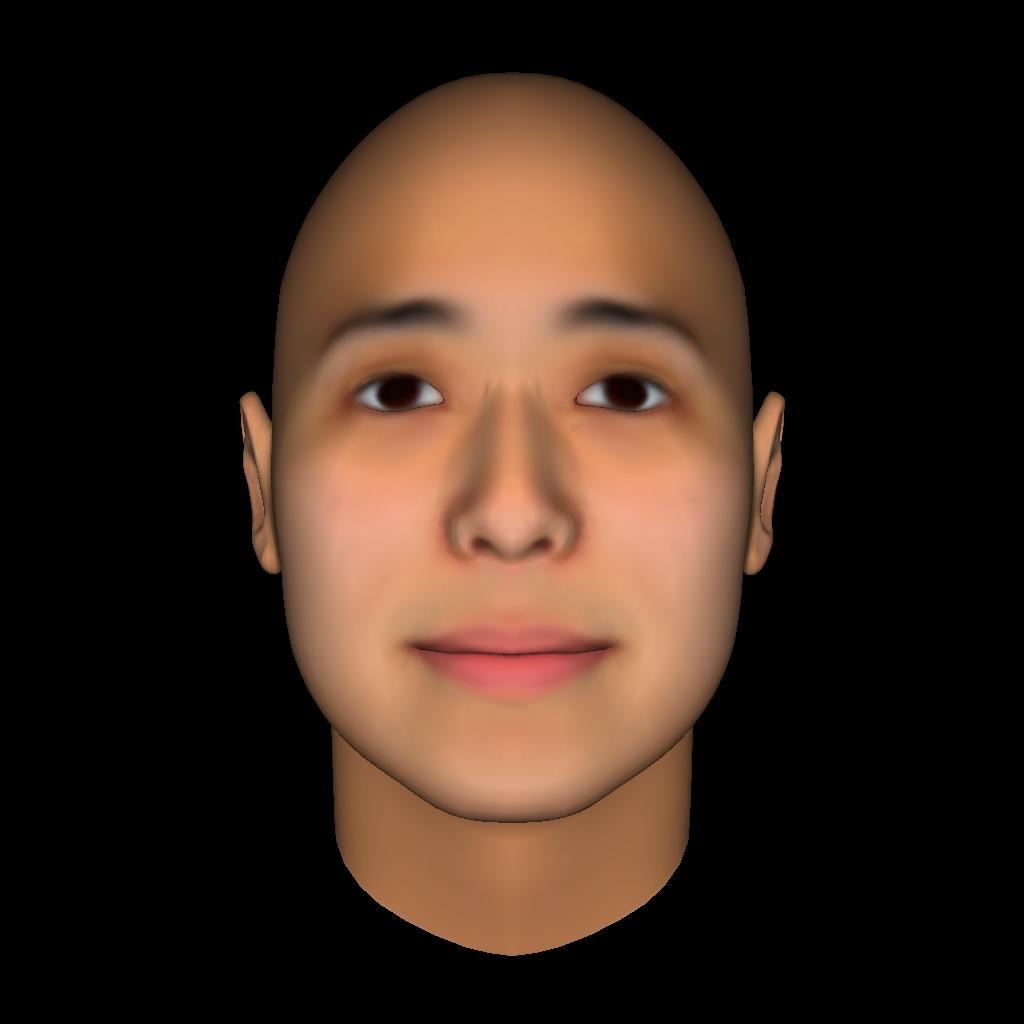}
    \includegraphics[width=0.2\linewidth, height=0.2\linewidth]{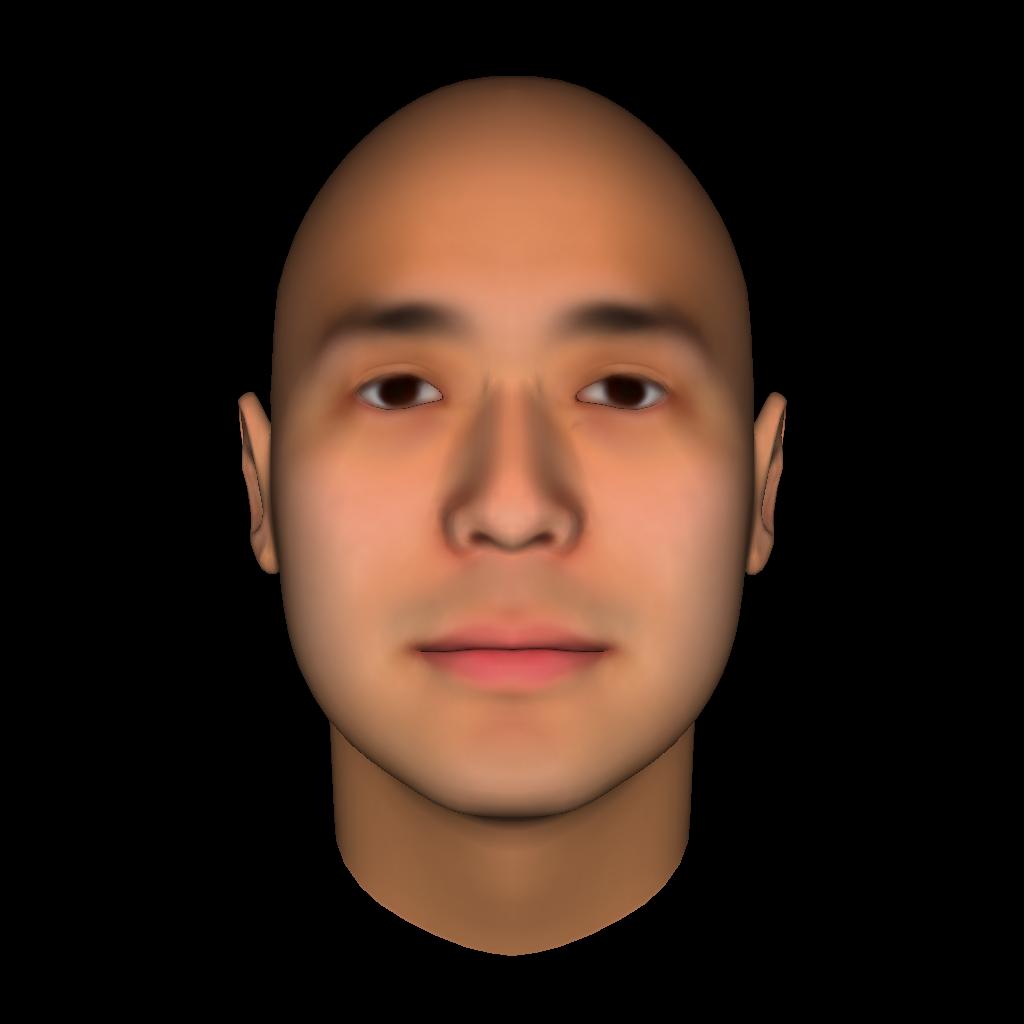}
    \includegraphics[width=0.2\linewidth, height=0.2\linewidth]{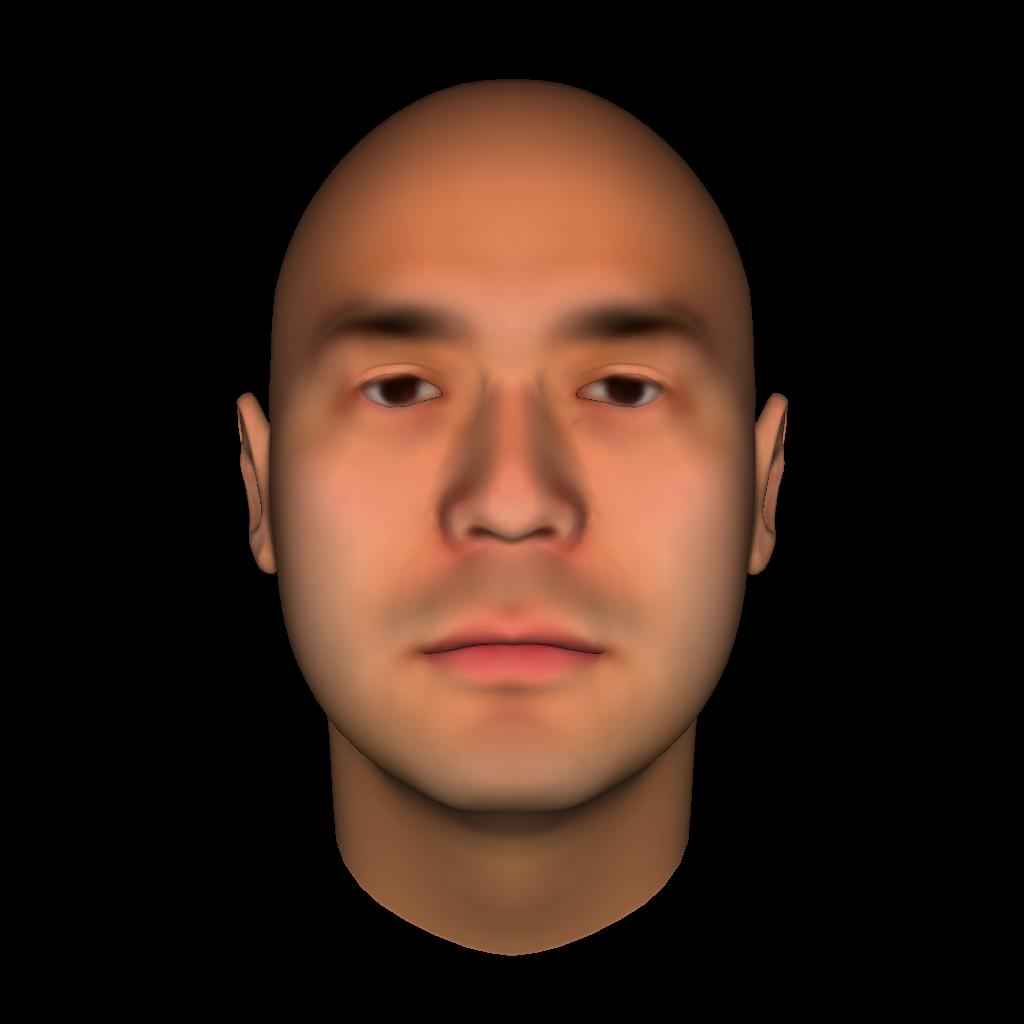}
    \\
    \includegraphics[width=0.2\linewidth, height=0.2\linewidth]{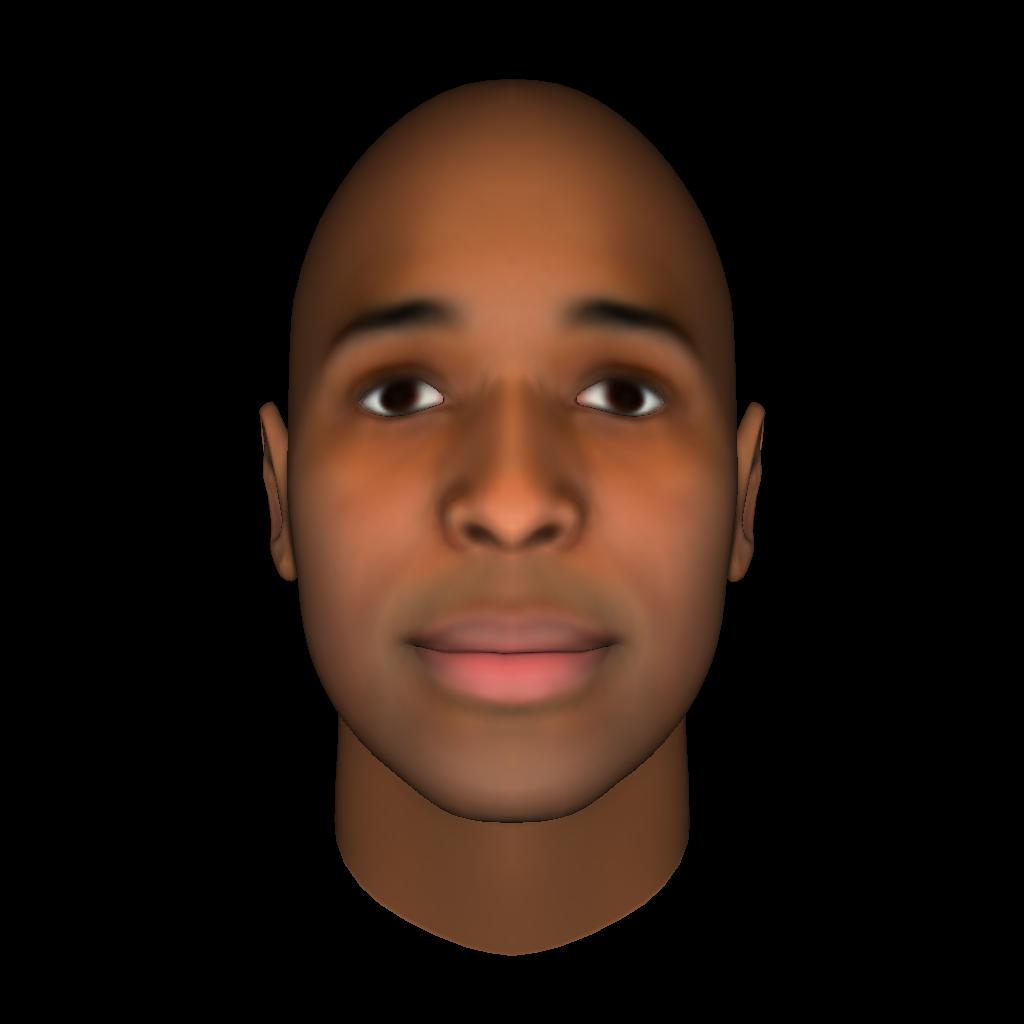}
    \includegraphics[width=0.2\linewidth, height=0.2\linewidth]{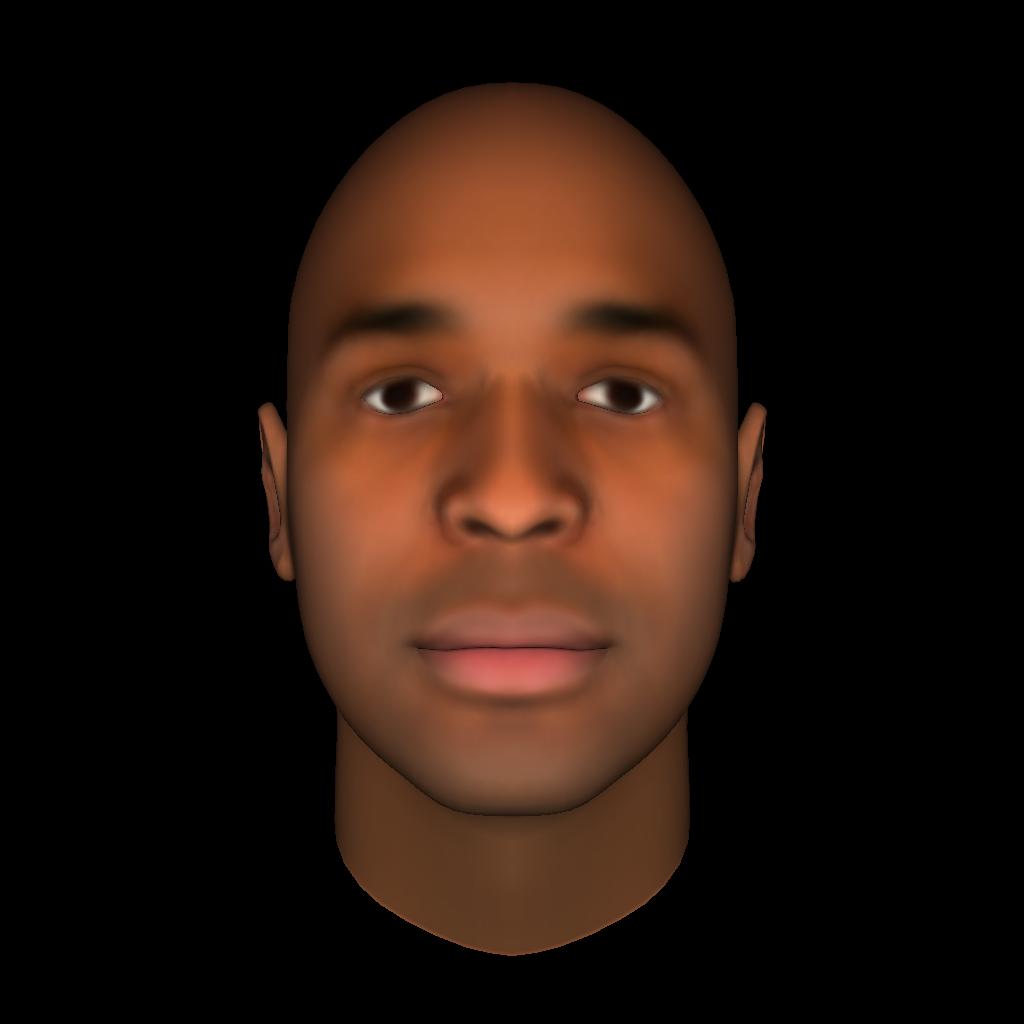}
    \includegraphics[width=0.2\linewidth, height=0.2\linewidth]{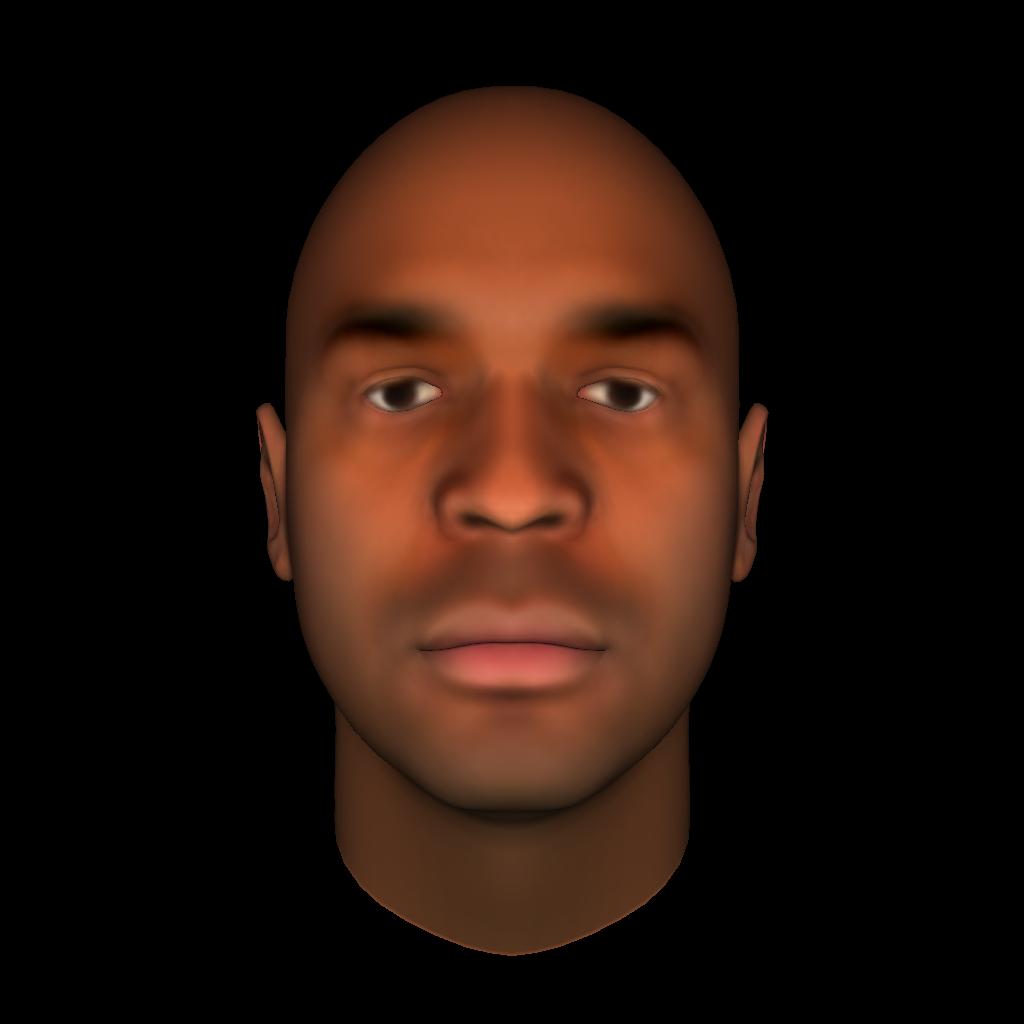}
    \caption{Faces (center) manipulated to appear 3SD more (left) and 3SD less (right) trustworthy than the average face \citep{Todorov2013ValidationFacial}.}
    \label{fig:face_distinct}
\end{figure}

To test whether first impression trait inferences can be learned from facial cues like the ones in Figure~\ref{fig:face_distinct}, we experiment with datasets of computer-generated faces developed to represent appearance bias in two psychological studies (\revision{R2P1}{Table~\ref{tab:datasets}}) \citep{Oosterhof2008TheEvaluation, Todorov2013ValidationFacial}. \revision{R1P1}{All the datasets used in our experiments can be obtained from the original authors at \url{http://tlab.princeton.edu/databases}}. These data come from a series of studies in which \citet{Todorov2013ValidationFacial} argue that computational models are the best tools for identifying the source of first impressions of facial features. In each study, human participants are shown a face for less than a second and then asked to rate the degree to which it exhibits a given trait (trustworthiness, competence, etc.) on a 9-point scale. Each face has a neutral expression, is hairless and is centered on a black background. The faces were generated with FaceGen, which uses a database of laser-scanned male and female human faces to create new, unique faces. \footnote{https://facegen.com/}

\begin{table*}[!t]

\begin{minipage}{\linewidth}
    \centering
    \caption{Sets of computer-generated faces with subjective human trait judgments}
    
\begin{tabular}{@{}p{10em} p{12em} p{4.5em} p{8em} l@{}}
\toprule
Name                     & Description                                                                                        & Race(s)                      & \# Faces                                     & Source \\ \midrule
\textsc{300 Random Faces}         & Randomly generated faces.                                                                          & Caucasian                    & 300                                          & \citep{Oosterhof2008TheEvaluation, Todorov2013ValidationFacial}   \\
\textsc{Maximally Distinct Faces} & Faces manipulated to exhibit a certain personality trait, according to subjective human judgments. & Caucasian, East Asian, Black & \makecell[{t}{p{8em}}]{1,875\\ (5 traits x 5 degrees x 75 identities)} & \citep{Oosterhof2008TheEvaluation, Todorov2013ValidationFacial}   \\
\textsc{Politicians}              & Faces of U.S. Senate, House, and Gubernatorial candidates, 1995-2008.                              & Any                          & \makecell[{t}{p{8em}}]{543\\ (246 Gubernatorial, 297 Senate)}          & \citep{Todorov2005InferencesOutcomes, Ballew2007PredictingJudgments}   \\ \bottomrule
\end{tabular}

    \label{tab:datasets}
\end{minipage}

\end{table*}

Together, these two sets provide a labeled benchmark for first impression, appearance-based evaluations of personality traits by human participants. One drawback to this training set is that because the computer-generated faces do not have hair and other accessories like make-up and glasses, our results may not generalize to images of real people on the web. Unfortunately, there are no large, publicly available datasets of experimentally validated trait judgments of real faces.

\subsection{Randomly Generated Faces}
The first dataset (\textsc{300 Random Faces}) includes 300 computer-generated, emotionally-neutral Caucasian faces created with FaceGen, a face generation software (Figure \ref{fig:face_distinct}). Though the face structures are gender-neutral, participants may still perceive bald faces as male \citep{Todorov2013ValidationFacial}. In a controlled laboratory setting, \citet{Todorov2013ValidationFacial} asked 75 Princeton University undergraduates to judge each face from this dataset on attractiveness, competence, extroversion, dominance, likeability, and trustworthiness \citep{Oosterhof2008TheEvaluation, Todorov2011Data-drivenPerception}. Here, the ground-truth labels are the trait scores provided by the study participants. \revision{R2P2}{Ideally, we would train on ground-truth labels for a larger number of randomly generated faces and for non-Caucasian faces, but none are available. Using only 300 randomly generated faces and 75 base faces may limit generalization to different types of faces. We leave additional data collection to future work.}

\subsection{Faces Manipulated Along Trait Dimensions}
For the second dataset (\textsc{Maximally Distinct Faces}), \citet{Todorov2013ValidationFacial} select 75 ``maximally distinct" faces from a random sample of 1,000 randomly generated Caucasian, East Asian, and Black faces. From this random sample of base faces, additional faces with maximally distinct perceived appearance bias were constructed as follows: using principal components analysis, the authors reduced the 3D FaceGen polygonal model that represents each base face to a 50-dimensional Euclidean vector space. Specifically, each component in the shape vector corresponds to a linear change in the positions of the vertices that structure the face \citep{Oosterhof2008TheEvaluation}. \citet{Oosterhof2008TheEvaluation} then find the best linear fit of the mean empirical trait judgments from \textsc{300 Random Faces} as a function of this shape vector. If $F \in \mathbb{R}^{50 \times 300}$ is a matrix of the shape vectors representing \textsc{300 Random Faces} with trustworthiness judgments $r\in\mathbb{R}^{300}$, then the optimal trustworthy vector is simply $t=F\cdot r$. Then a face with shape vector $\alpha$ can be manipulated to appear $\delta$ SD more trustworthy with the new vector $\alpha' = \alpha + \delta \cdot \hat{t}$, where $\hat{t}$ is the normalized trustworthiness gradient vector. This method leverages the ground-truth subjective trait judgments to compute the optimal direction in which to alter the subjective trustworthiness - or another trait - of a randomly generated face in FaceGen.

\citet{Todorov2013ValidationFacial} use this method to generate faces that vary along each trait dimension to produce a set of faces to elicit a trait inference -3, -2, -1, 0, 1, 2, and 3 SD from the mean - 25 variations in total for each of the 75 faces, resulting in a total 1,875 labelled faces for each trait (Figure~\ref{fig:face_distinct}). These manipulations are not necessarily related to facial expression \citep{Oosterhof2008TheEvaluation}. Though the perturbations themselves are not psychologically meaningful and do not deliberately correspond to any particular facial features or expressions, these manipulations tend to produce faces that vary noticeably along the trait dimensions (Figure \ref{fig:face_distinct}). Each face was presented to 15 different Princeton university students for subjective scoring on the same 9-point scale used in \textsc{Random Faces}. These scores were validated for interrater reliability (using Cronbach's $\alpha$ for all average trait ratings) and explained variance when regressed on the standard deviation scores targeted by the face-generation model; studies with human participants confirm that the manipulated faces do on average alter the subjective appearance of a given face by $\delta$ SD \citep{Todorov2013ValidationFacial}. Faces produced with the maximally distinct method are reliable indicators of human trait judgments. 

Since the authors show that there is a high degree of correlation between the average human trait score and the target SD, and validation scores are not available for every image in the dataset, we use the target SD scores as labels for training.  So that the ground-truth labels for both \textsc{Random Faces} and \textsc{Maximally Distinct Faces} are scaled identically and can be trained on simultaneously, we convert the raw 9-point scale used for the \textsc{Random Faces} labels to standard (z-) scores such that both sets of labels are measured in terms of standard deviation from the mean.

\subsection{Real Faces}
We also validate our model on a small set of real (not computer-generated) faces. The \textsc{Politicians} dataset includes the faces of U.S. Senate, House, and Gubernatorial candidates from 1995 to 2008, evaluated by human participants on the basis of apparent competence \citep{Todorov2005InferencesOutcomes}. \citet{Todorov2005InferencesOutcomes} and \citet{Ballew2007PredictingJudgments} show that 1-second inferences of politicians' competence based only on facial appearance are linearly related to their margin of victory. Participants were presented with the winner and runner-up of each election and asked to judge which person was more competent (without knowing the result), in binary and on the same 9-point scale used in \textsc{300 Random Faces}. Participants were more likely to choose the winner  \citep{Todorov2005InferencesOutcomes, Ballew2007PredictingJudgments}. Other trait judgments are also included in the dataset, but there is no significant effect for traits other than competence \citep{Todorov2005InferencesOutcomes}.

\section{Approach}
\label{sec:approach}

\begin{figure}[!t]
    \centering
    \includegraphics[width=\linewidth]{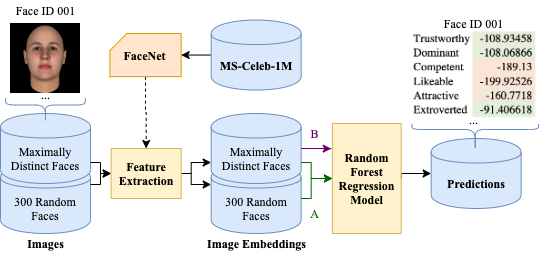}
    \caption{A transfer learning model trained on subjective trait scores. FaceNet, pre-trained on the MS-Celeb-1M benchmark dataset, extracts embeddings for each face. In \textbf{Experiment A}, a random forest regression model is trained on feature embeddings from the set of faces manipulated to be maximally distinct and the set of randomly generated faces with human scores. \textbf{Experiment B} compares these two sets of training images with a regression trained only on the randomly generated faces.}
    \label{fig:model}
\end{figure}

We construct a transfer learning model to leverage face representations extracted from a pre-trained, state-of-the-art face recognition model (Figure \ref{fig:model}). Since we are testing whether standard industry methods are capable of learning trait inferences, we use popular industry techniques for pre-processing and modeling. First, according to the state-of-the-art, we crop and align every face with pose estimation to ensure the faces have similar size, shape, and rotation \citep{kazemi2014one}. By cropping out the bald head, we also make the computer-generated images seem more gender-neutral. Then, from the final layer of FaceNet, a popular open-source Inception-ResNet-V1 deep learning architecture, we extract a standard 128-dimensional feature vector from the pixels of each transformed image \citep{Schroff2015FaceNet:Clustering}. For thousands of images, extraction takes minutes. Rather than train FaceNet from scratch, we utilize a model with weights pre-trained on the MS-Celeb-1M dataset, a common face recognition benchmark \citep{Guo2016MS-Celeb-1M:Recognition}, downloaded from \url{https://github.com/davidsandberg/facenet}. We chose this transfer learning approach to mitigate the fact that our dataset is entirely computer-generated: by using a model pre-trained on real faces, we can extract features more similar to features from images in the wild. MS-Celeb-1M contains 10 million images of one million celebrities and was one of the largest publicly available face recognition benchmark datasets, making it a popular choice for transfer learning face recognition models, before Microsoft took it down in 2019 \citep{PearsonMicrosoftDead}. Notably, MS-Celeb-1M was reportedly used to train controversial mass surveillance algorithms in China \citep{MurgiaWhosTimes}. By using a pre-trained model for feature extraction, we imitate feature processing techniques used commonly in black box industry models. The FaceNet model (over 10 thousand stars on Github), and similar architectures such as OpenFace (over 13 thousand stars on Github), are used by software developers, researchers, and industry groups \citep{Schroff2015FaceNet:Clustering, Amos2016OpenFace:Applications}.

After feature extraction, we train six random forest regression models\footnote{We also tested an SVM model (with RBF kernel) and a logistic regression model (l2 penalty, $C=1.0$). The SVM performed about 3 points worse in cross validation than the logistic regression or random forest models. We chose the random forest model because 1) with bootstrap aggregating and random feature selection it tends not to overfit \cite{Breiman2001RandomForests}, and 2) parallelization for quicker execution.} to predict appearance bias for each of the six traits measured: attractiveness, competence, dominance, extroversion, likeability, and trustworthiness.
The human participants' trait scores, multiplied by 100 for readability, serve as the ground-truth labels. The random forest includes 100 weak learners with no maximum depth, a minimum split size of two, and mean-squared-error split criterion. Data and code used to produce the figures, tables, and pre-trained model (Figure~\ref{fig:model}) in this work are available at \github.

\section{Experiments \& Results}
\label{sec:results}

\subsection{Learning appearance bias}
We validate our model's ability to learn human appearance bias scores under several different experimental conditions. In general, we find that our model is capable of learning appearance bias from manipulated faces with a high degree of accuracy, but fails to make accurate predictions for randomly generated faces.

\paragraph{Experiment A} To test how well the random forest regression model learns appearance bias from both sets of labeled faces (randomly generated and computer-generated), we shuffle the image embeddings extracted with FaceNet such that the 300 random faces and maximally distinct faces are mixed. The target labels are the original appearance bias measurements provided by human participants. Splitting the training data into 10 equal folds, we do the following for each fold: 1) train the regressor on the other 9 partitions; 2) record and plot appearance bias predictions for the current partition. Once all 10 partitions are processed, each image has a corresponding vector of predicted appearance bias scores, one for each trait measured. Table~\ref{tab:coefs} displays goodness-of-fit and correlation statistics from the cross-validations for regressions on all six traits measured. For reference, \citet{Safra2020TrackingPaintings}, who train a model on a subset of \textsc{Maximally Distinct Faces} to predict trustworthy ratings, report significant correlation coefficients of $\rho=0.85$ for trustworthiness and $\rho=0.86$ for dominance for cross-validation on a held-out test set of maximally distinct faces. In contrast, with 10-fold cross validation, we achieve significant correlation coefficients of $\rho=0.98$ and $\rho=0.99$, respectively. Likely, the higher accuracy is a result of our larger training set (\citet{Safra2020TrackingPaintings} train on only the Caucasian maximally distinct faces).

\begin{figure}[!ht]
    \centering
    \begin{subfigure}[t]{0.48\textwidth}
        \centering
        \includegraphics[width=\textwidth]{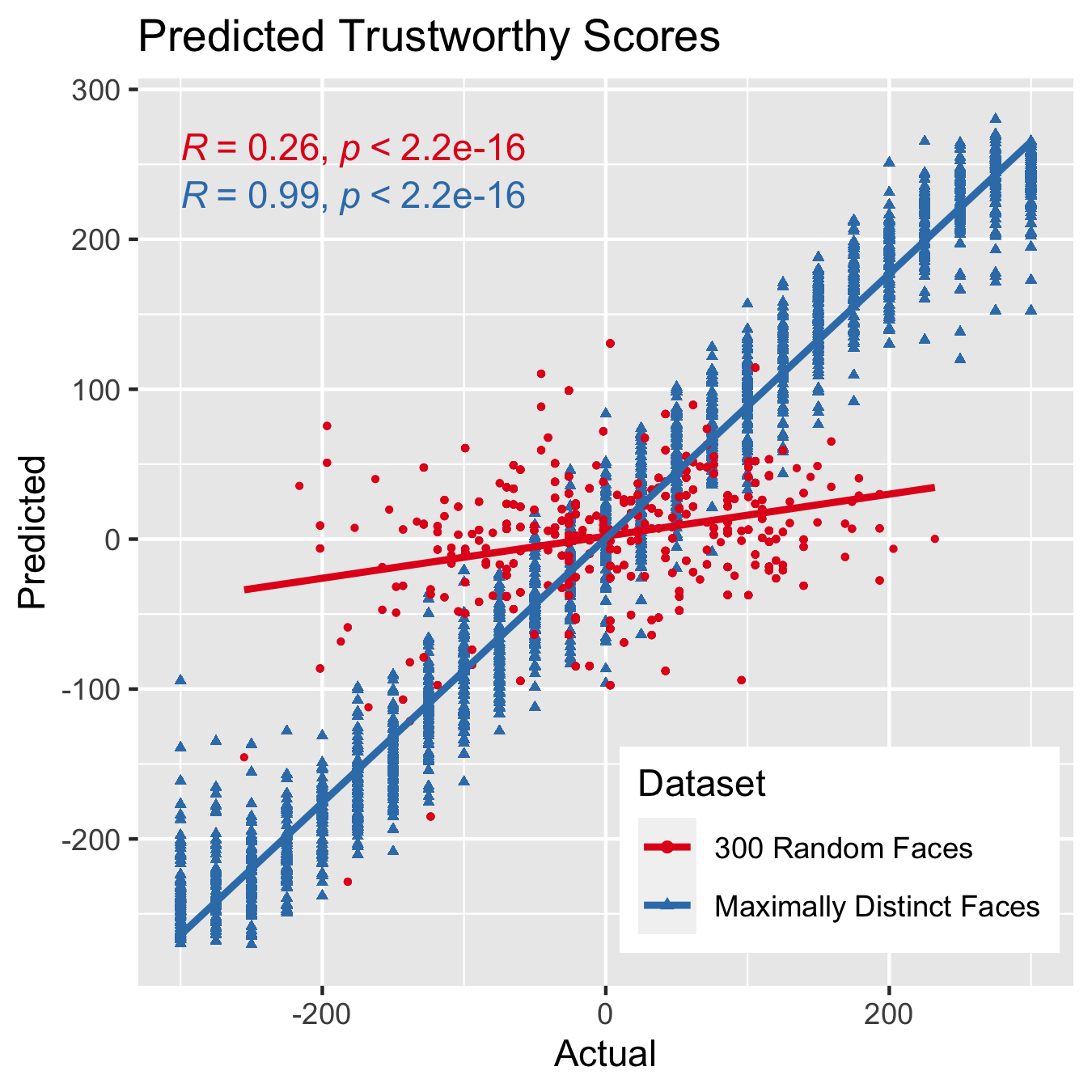}
        \caption{Model \textbf{A}, trained on both datasets.}
    \end{subfigure}
    \quad
    \begin{subfigure}[t]{0.48\textwidth}
        \centering
        \includegraphics[width=\textwidth]{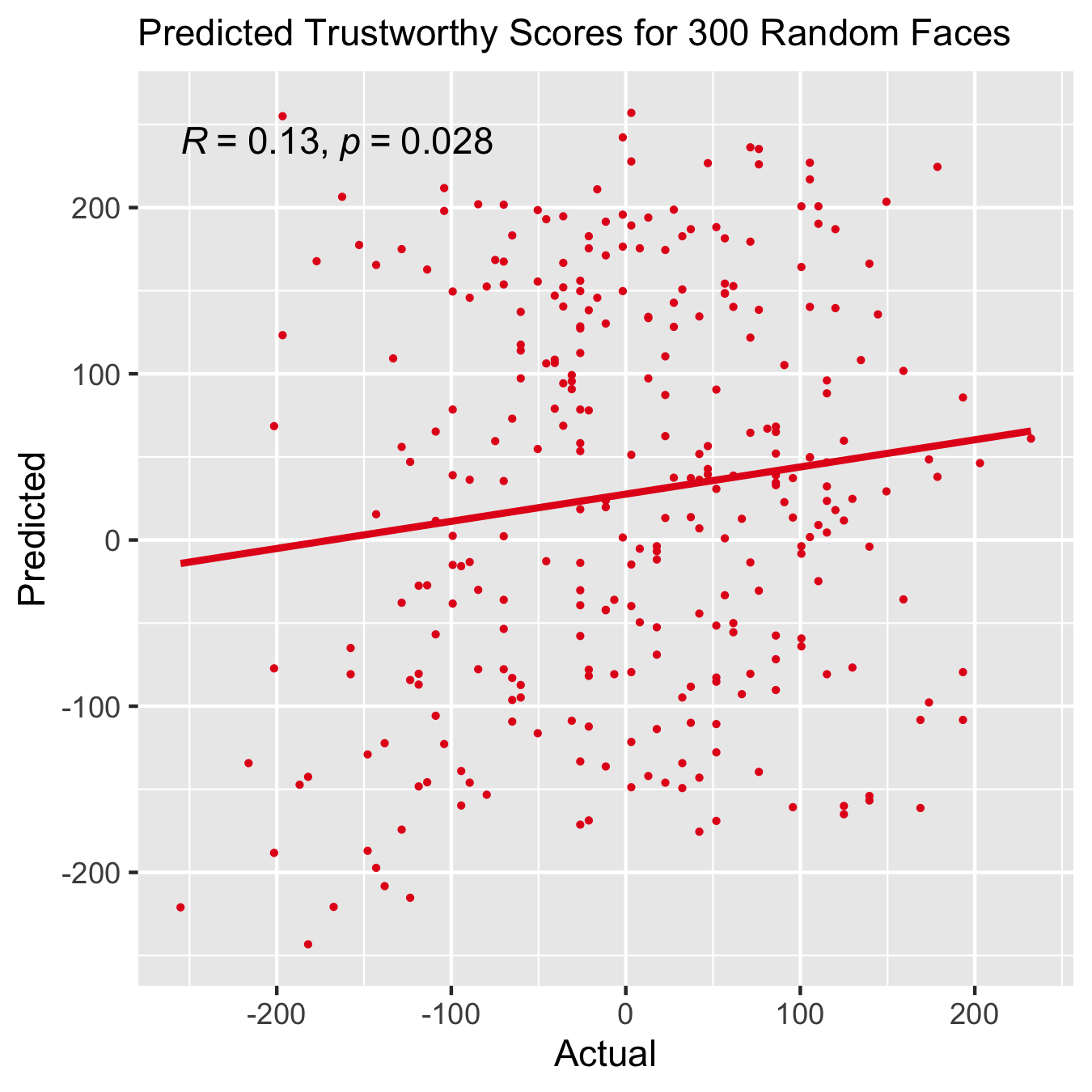}
        \caption{Model \textbf{B}, trained on only the maximally distinct faces and tested on the 300 random faces.}
    \end{subfigure}
    \caption{Fit line and scatter plot of actual ``trustworthiness" impressions against 10-fold cross-validated predictions for models trained on both sets of computer-generated faces (\textbf{A}) or just the randomly generated faces (\textbf{B}).}
    \label{fig:scatter}
\end{figure}

Notably, our approach learns appearance bias to a high degree of precision for the maximally distinct faces ($\rho=.99)$, but the accuracy drops on when predicting human trait scores for randomly generated faces (Figure~\ref{fig:scatter}). The model performs poorly on randomly generated faces even when randomly generated faces are included in the training set, suggesting there is either no consistent signal in trait scores of random faces, that the signals are too complex to be learned by this model, or that the transfer learned face representations used for training do not contain useful information for predicting trait inferences. The former explanation seems unlikely; human participants tended to agree on trait scores for the random faces:  the interrater reliability for \textsc{Random Faces} was $\alpha=0.84$, roughly the same as for \textsc{Maximally Distinct Faces}. In other words, human participants tended to predict the judgments of other human participants, so there is some signal to be modelled). For the remainder of the paper, we will explore the second two explanations for the low correlation between our model's predictions and human appearance biases.

\paragraph{Experiment B} To better assess our model's performance and investigate the disparity in predictive performance on the maximally distinct faces and the randomly generated faces, we train the regression model on only the maximally distinct faces and test on only the randomly generated faces. Though a 10-fold cross validation of the maximally distinct model has an average explained variance of 97\% and an average prediction correlation of 99\%, prediction on the randomly generated faces is much less accurate than in Experiment A ($\bar{\rho}=.32)$.\footnote{For all traits, the errors are distributed approximately normally.} Like the human participants, our model tends to agree more about judgments of deliberately manipulated faces than about judgments of randomly generated faces. Our approach learns subjective scores of appearance bias, generated in a controlled experiment, more accurately with respect to judgments of dominance than judgments of other traits, perhaps because dominance has been shown to be less correlated with facial cues than other traits \citep{Willis2006FirstImpressions}. The standard deviation in dominance scores on the original 9-point scale is 1.14, much higher than the average 0.72 standard deviation for the other traits. 

For reference, \citet{Safra2020TrackingPaintings} report significant correlation coefficients of $\rho=0.22$ for trustworthiness and $\rho=0.16$ for dominance when validating on 4 external databases of real faces with human-annotated bias scores. Compared to the high correlation scores for the computer-generated faces, both our results and those of \citet{Safra2020TrackingPaintings} convey a fairly large generalization gap, suggesting both models struggle to generalize to non-maximally distinct faces. \revision{R1P3}{Some of this gap may be attributed to differences in the two separate groups of study participants from which the two sets of ground-truth labels were sourced, but like \citet{Todorov2013ValidationFacial}, we assume that the participants were selected from the same population and that there is no systematic difference in the biases of the two groups. }That our model's performance drops significantly for randomly generated faces suggests the generalization gap is due not only to the uncanny differences between computer-generated faces and real faces but also to overfitting on the particular feature dimensions that were manipulated during maximally distinct face generation.

\begin{table*}[!t]

\begin{minipage}{\linewidth}
    \centering
    \caption{Correlation of actual and predicted appearance biases}
    \begin{tabular}{@{} r l l l l l l l @{}} 
    \toprule
    \# &
    Traits & Attractive & Competent & Dominant &  Extroverted &    Likeable & Trust \\ \midrule
    \multirow{3}{*}{A} &  $\rho$ &     0.99  & 0.99 &  0.99 &  0.98  & 0.99  & 0.98 \\ %
    & p-value &       $<10^{-16}$ &  $<10^{-16}$ &  $<10^{-16}$ &  $<10^{-16}$  & $<10^{-16}$ &  $<10^{-16}$ \\ %
    & RMSE & 30.3 & 33.3 & 27.6 & 36.6 & 33.4 & 35.9 \\ \hline
    \multirow{3}{*}{B} &  $\rho$&    0.30 &  0.29 &  0.72  & 0.25  & 0.17 &  0.13\\ %
    & p-value &   $<10^{-6}$  &  $<10^{-6}$  &  $<10^{-16}$ &   $<10^{-4}$  &  $<10^{-2}$ &   $0.028$\\ %
    & RMSE & 134.2 & 152.2 & 141.0 & 105.9 & 134.5 & 144.8 \\ \bottomrule
\end{tabular}
    \caption*{Pearson's correlation coefficient $\rho$ and root mean square error (RMSE) for regression predictions. In Experiment \textbf{A}, a random forest regression is fitted on both sets of faces and predictions are produced by 10-fold cross validation; in \textbf{B}, the regression is fitted on maximally distinct faces and tested on randomly generated faces. P-values are from the correlation t-test of $H_0: \rho=0$.}
    \label{tab:coefs}
\end{minipage}

\end{table*}

\paragraph{Experiment C} Do these results hold if the problem of learning appearance bias is treated as a classification problem? We binarize the ground-truth trait judgments into ``positive" and ``negative" classes (e.g. ``Trustworthy" and ``Not Trustworthy") and train a random forest classifier, instead of a random forest regressor, on the class-labeled face embeddings. Again, the model performs well when tested on maximally distinct faces, with 95\% accuracy for the trustworthy trait in 10-fold cross-validation, but poorly when tested on random faces (46\% accuracy). If the model is trained only on the random faces, it achieves a 10-fold cross-validation accuracy of only 43\%. For both models, false negatives and false positives occur at roughly the same rate.

\paragraph{Experiment D} Though the model performs no better than chance on the randomly generated faces, perhaps the bias judgments learned from the maximally distinct faces will predict bias in real, human faces. We train a model on the computer-generated \textsc{Maximally Distinct Faces} dataset to predict competence scores for the \textsc{Politicians} faces. There is no significant correlation between the predicted competence scores and the competence scores collected from human participants, according to a Pearson's product-moment correlation t-test. The RMSE for this model is 98.8, significantly higher than in Experiment B but much worse than in Experiment A. Further, \citet{Ballew2007PredictingJudgments} find that competency judgments predict 2006 Gubernatorial and Senator election winners at an average rate of 68.6\% ($p<0.008$) and 72.4\% ($p<0.016$) against chance, respectively, according to a 1-sample chi-square test of proportion. Our predicted competency judgments only predict winners at an average rate of 45.7\% ($p=0.61$) in Gubernatorial races and 67.9\% ($p < 0.1$) in Senate races. For comparison, random chance would predict the correct winner 50\% of the time. Neither result differs significantly from chance at more than 95\% confidence. There is no significant correlation between predicted competence and vote difference ($p=0.20$), but there is a slight correlation ($\rho=0.21,\;p < 10^{-3}$) between the difference in predicted competence scores between two candidates in a race and the vote spread. In summary, a model trained on random faces and maximally distinct faces also fails to generalize to real-world faces. \revision{R2P3}{Perhaps a model trained on more randomly generated faces would generalize better than a model trained solely on maximally distinct faces, but there are not enough ground-truth labels. We leave this to future work.}

\subsection{Feature Analysis}
\label{subsec:feature-analysis}
Why does our model perform well on the maximally distinct faces, but poorly in the wild? Equally poor performance on the computer-generated random faces (Experiment B) suggests that generalization from computer-generated faces to real faces is not the only challenge in learning appearance bias.

\begin{figure}[!ht]
    \centering
    \begin{subfigure}[t]{0.48\textwidth}
        \centering
        \includegraphics[width=\textwidth]{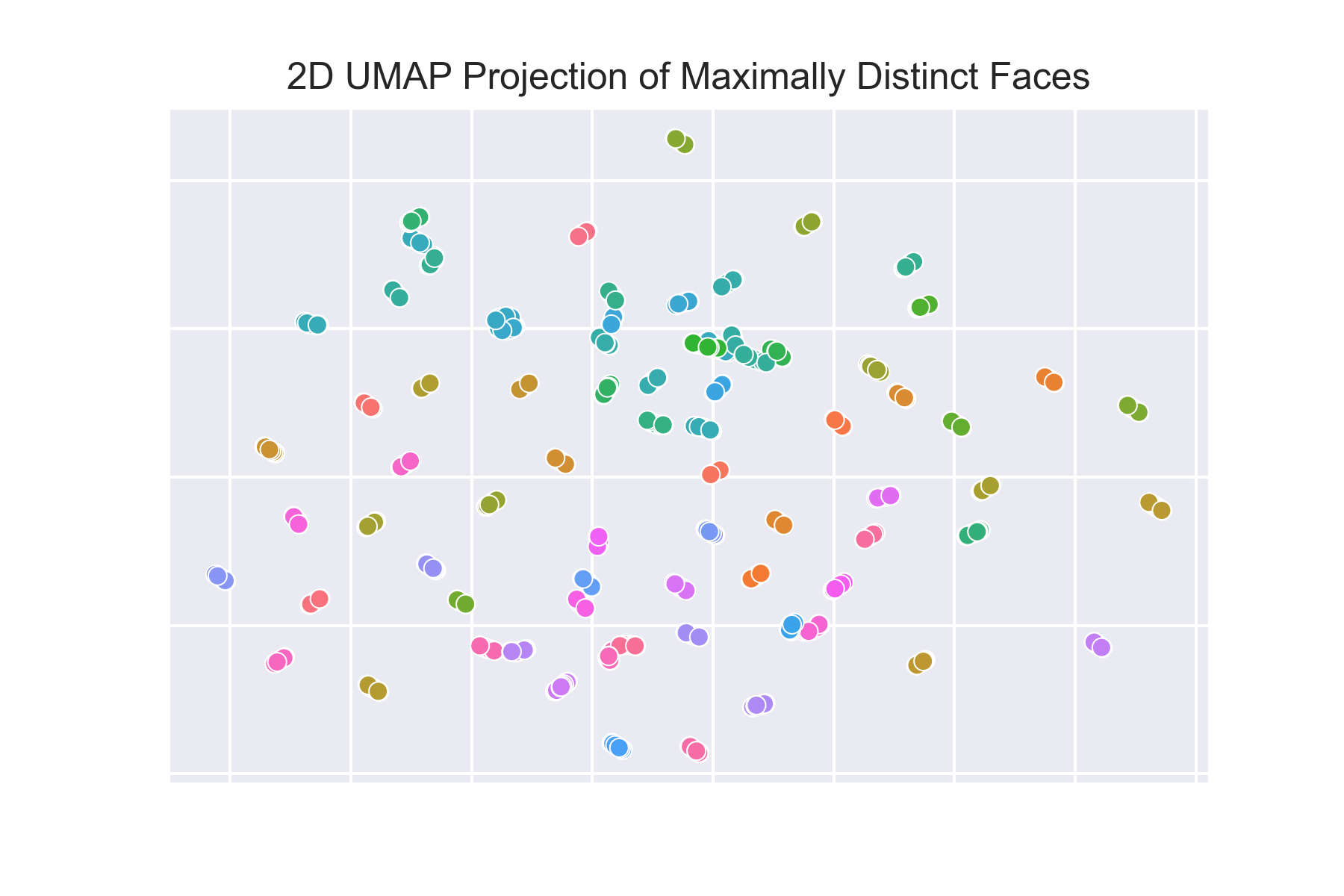}
        \caption{Features from \textsc{Maximally Distinct Faces}, shaded by face identity (which randomly generated face was manipulated to produce this face). UMAP tends to cluster maximally distinct faces that were created by manipulating the same base face.}
    \end{subfigure}
    \quad
    \begin{subfigure}[t]{0.48\textwidth}
        \centering
        \includegraphics[width=\textwidth]{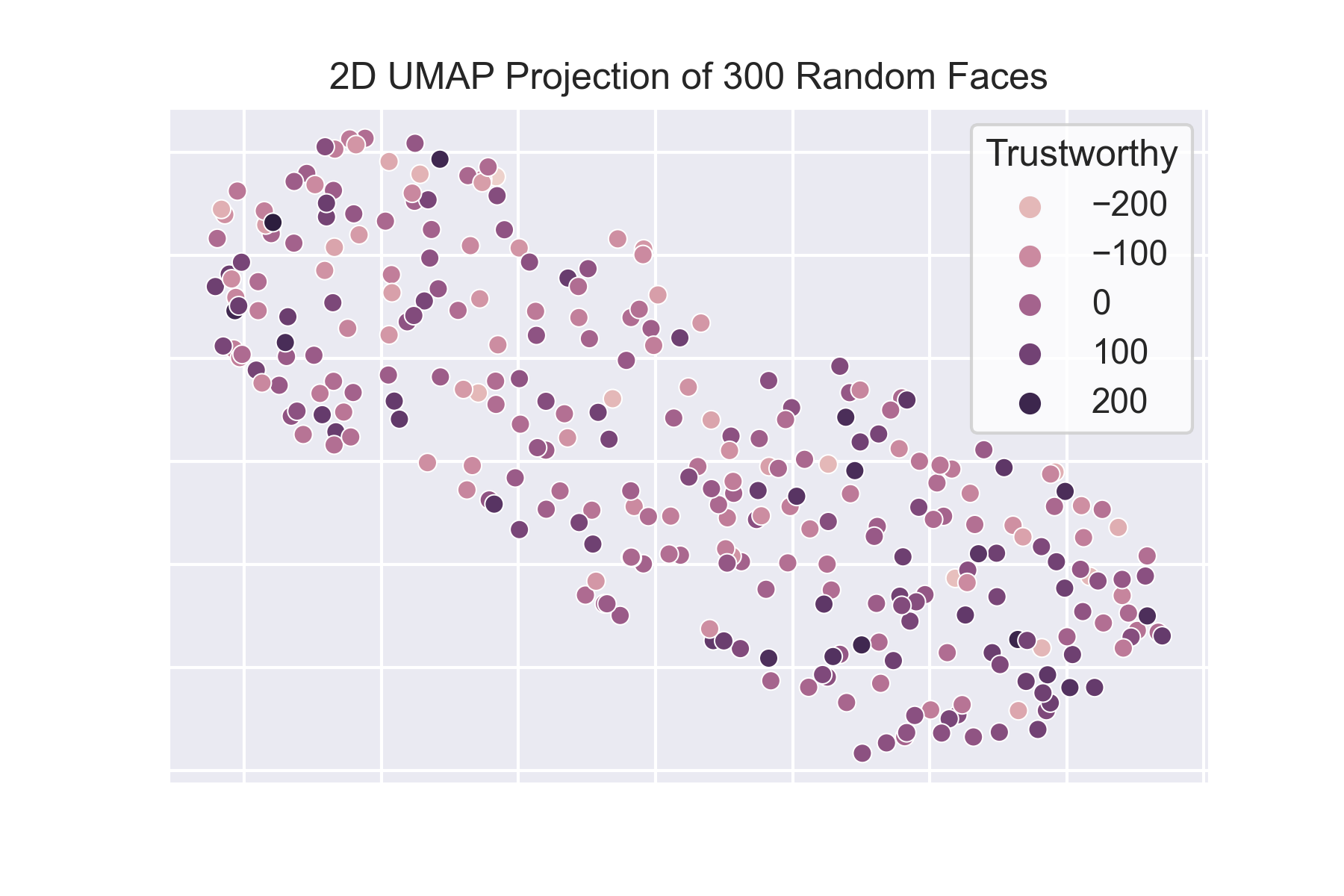}
        \caption{Features from \textsc{300 Random Faces}, shaded by perceived trustworthiness (in SD from the mean, multiplied by 100).}
    \end{subfigure}
    \caption{2D UMAP projection of face features for both datasets with 15 features, minimum distance of 0.1, and 2 components. We use a high number of features to avoid spurious clustering \citep{McInnes2018UMAP:Reduction}.}
    \label{fig:umap}
\end{figure}

\paragraph{Face embeddings} Though FaceNet embeddings clearly differentiate each individual face in the dataset, they are not designed to represent the facial features relevant to trait judgments. Using Uniform Manifold Approximation and Projection (UMAP), we cluster embeddings of both the \textsc{Maximally Distinct Faces} and the \textsc{Random Faces} (Figure~\ref{fig:umap}). UMAP is a popular unsupervised clustering algorithm for image data, good for efficiently capturing the global structure of high-dimensional data \citep{McInnes2018UMAP:Reduction}. Recall that the maximally distinct faces are created by manipulating \revision{R1P4}{each of } 75 random faces into 175 different faces spread along a trait dimension. In 3D space, each maximally distinct face tends to be clustered with its manipulated siblings despite variation across the trait axis. Likewise, there is no pattern of trait clustering in the distribution of the random faces. Despite computer manipulation of trait appearances in the maximally distinct faces, an unsupervised projection of FaceNet embeddings emphasizes differences between individual faces, not differences in features that contribute to appearance biases. \revision{R1P4}{The unsupervised industry-standard face embedding model we use in this study is designed to embed features that distinguish individual faces in a variety of poses and expressions, allowing face recognition classifiers trained on these embeddings to more easily generalize to new settings. But evidently, these unsupervised embeddings do not automatically distinguish faces according to subjective traits. As a result, the final, supervised classifier struggles to generalize to real-world datasets.}

\begin{figure}[!ht]
    \centering
    \begin{subfigure}[t]{0.45\textwidth}
        \centering
        \includegraphics[width=\textwidth]{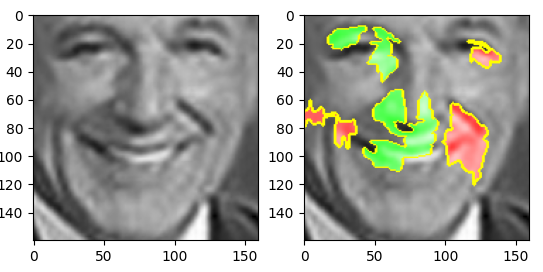}
        \caption{
            Face with greatest prediction error. From 1996 Gubernatorial race, UT.\vspace{1em}\\
            \emph{Predicted competence:} \hfill+2.27SD \\\emph{Ground-truth competence:} \hfill-1.61SD
        }
    \end{subfigure}
    \quad
    \begin{subfigure}[t]{0.45\textwidth}
        \centering
        \includegraphics[width=\textwidth]{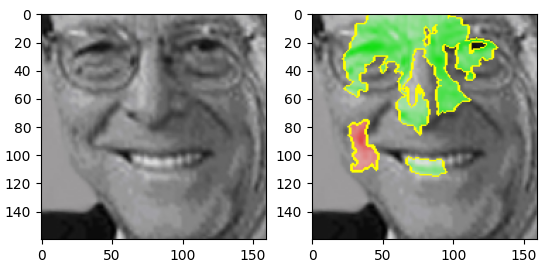}
        \caption{
            Face with least prediction error. From 2002 Senate race, NM.\vspace{1em}\\
            \emph{Predicted competence:} \hfill+0.72SD \\\emph{Ground-truth competence:} \hfill+0.72SD
        }
    \end{subfigure}
    \caption{LIME explanations for predicted competence judgments of politician faces. The 10 most important superpixels are shaded. Green indicates agreement with the ultimate prediction; red indicates disagreement.}
\label{fig:lime-pol}
\end{figure}

\paragraph{Feature importance} What facial features is our model using to make trait judgments? We generated Local Interpretable Model-Agnostic Explanations (LIME) for each face in \textsc{Random Faces} and \textsc{Politicians} \citep{Ribeiro2016WhyClassifier}. LIME is a popular black box interpretability tool that approximates an interpretable local model for the classification version of this problem (Experiment C). Taking our model and a test sample as input, LIME perturbs the superpixels of the sample face and measures the corresponding changes in our model's prediction. These changes indicate which groups of pixels or most important to our model's prediction and whether they agree with or contradict the final prediction. Images are segmented into 300 superpixels with Simple Linear Iterative Clustering (SLIC), enough to capture facial features as small as the pupil \citep{Achanta2012SLICMethods}. We generate explanations in a neighborhood of 5,000 samples; large samples tend to reduce variability in the outputted feature weights \citep{Ribeiro2016WhyClassifier}. Figure~\ref{fig:lime-pol} depicts two example explanations for the greatest and least prediction errors in the \textsc{Politicians} out-of-sample test set.

\begin{figure}[!ht]
    \centering
    \begin{subfigure}[t]{0.31\textwidth}
        \centering
        \includegraphics[width=0.8\textwidth]{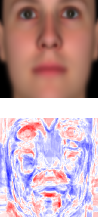}
        \caption{All faces.}
    \end{subfigure}
    \quad
    \begin{subfigure}[t]{0.31\textwidth}
        \centering
        \includegraphics[width=0.8\textwidth]{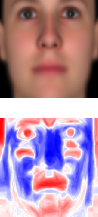}
        \caption{Predicted trustworthy.}
    \end{subfigure}
    \quad
    \begin{subfigure}[t]{0.31\textwidth}
        \centering
        \includegraphics[width=0.8\textwidth]{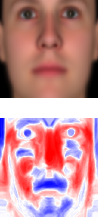}
        \caption{Predicted untrustworthy.}
    \end{subfigure}
    \caption{Average face (top) and corresponding heatmap of average LIME explanation across all predicted trustworthy judgments of \textsc{300 Random Faces}. Blue indicates agreement with the ultimate prediction; red indicates disagreement.}
\label{fig:lime-agg-rand}
\end{figure}

\begin{figure}[!ht]
    \centering
    \begin{subfigure}[t]{0.31\textwidth}
        \centering
        \includegraphics[width=0.8\textwidth]{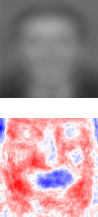}
        \caption{All faces.}
    \end{subfigure}
    \quad
    \begin{subfigure}[t]{0.31\textwidth}
        \centering
        \includegraphics[width=0.8\textwidth]{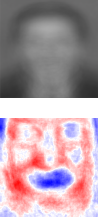}
        \caption{Predicted trustworthy.}
    \end{subfigure}
    \quad
    \begin{subfigure}[t]{0.31\textwidth}
        \centering
        \includegraphics[width=0.8\textwidth]{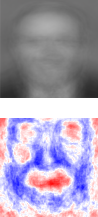}
        \caption{Predicted untrustworthy.}
    \end{subfigure}
    \caption{Average face (top) and corresponding heatmap of average LIME explanation across all predicted competence judgments of \textsc{Politicians}. Blue indicates agreement with the ultimate prediction; red indicates disagreement.}
\label{fig:lime-agg-pol}
\end{figure}

In the \textsc{300 Random Faces} dataset, the features which contribute most (positively or negatively) to the final prediction are consistently clustered around the eyes, nose cheekbones, mouth, and upper lip \revision{R2P5}{(Figure~\ref{fig:lime-agg-rand})}. Occasionally, particularly for lighter faces, features are scattered all across the face, but usually not in the background. \revision{R2P5}{These observations hold for photos of real people: though there is additional variance in face position, features clustered around the average position of the mouth, cheekbones, and eyes contribute most to the final competence prediction for \textsc{Politicians} dataset (Figure~\ref{fig:lime-agg-pol}). }There do not appear to be any differences in allotment of feature importance between trustworthiness, competence, and other traits. This result may be surprising, but the models used to generate the training data (\textsc{Maximally Distinct Faces}) do not explicitly manipulate particular facial features \citep{Oosterhof2008TheEvaluation}. \revision{R2P5}{In both datasets, our model appears to rely on the same facial features to classify faces for multiple traits; according to our results, these features are not predictive of human appearance biases.}

\section{Discussion \& Conclusions} 
\label{sec:discussion}

Though our model can learn appearance bias from a small set of maximally distinct, computer manipulated faces, it fails to make similar trait judgments out-of-sample and does not exhibit the same biases in the wild as people do. This result casts doubt on the use of computationally manipulated features to learn appearance bias: our model is trained on the same data as \citet{Safra2020TrackingPaintings}, who achieve similarly low correlation scores on out-of-sample faces. With clustering and interpretability analyses, we identify two explanations for this phenomenon. First, there is insufficient overlap between state-of-the-art embeddings for face recognition and the features required to identify appearance biases in real and random faces, if they exist at all. \revision{R2P5}{Second, though 1) the trait dimensions identified and manipulated by \citet{Todorov2013ValidationFacial} to produce maximally distinct faces match human appearance biases and 2) similar features can be used to explain our model's predictions, these features are not predictive of subjective human judgments for real or even randomly generated faces. In short, industry-standard face recognition is not sufficient to learn subjective human judgments from this computational model of trait perception.}

If, as \citet{Todorov2013ValidationFacial} claim, ``computational models are the best available tools for identifying the source of [trait] impressions," then more research is needed to construct externally valid representations of appearance bias. \revision{R1P2}{For example, the ground-truth measures of subjective trait judgments currently available are sourced from largely white, young Princeton students, whose appearance biases are not globally representative. } Further, predictions from transfer learning models trained on maximally distinct, computer-generated features provide neither an objective measure of personality traits (they represent subjective biases) nor a good measure of subjective bias itself, as we show. However, our results do not rule out the possibility that appearance bias could be embedded from a larger training set of real faces with labels from a more representative set of participants: future work should investigate whether human appearance bias manifests in large-scale datasets in the wild.

\section*{Acknowledgements}
We thank two anonymous reviewers for their insightful comments and helpful feedback; and in particular, we thank one anonymous reviewer for inspiring Figures \ref{fig:lime-agg-rand} and \ref{fig:lime-agg-pol}.

This material is based on research partially supported by the U.S. National Institute of Standards and Technology (NIST) Grant 60NANB20D212. Any opinions,  findings, and conclusions or recommendations expressed in this material are those of the authors and do not necessarily reflect those of the NIST.

\bibliographystyle{plainnat}
\bibliography{references}

\end{document}